\documentclass[a4paper,11pt]{article}
\pdfoutput=1 

\usepackage{jinstpub} 

\usepackage{lineno}
\usepackage{subfig}
\usepackage{natbib}
\usepackage[utf8]{inputenc}

\title{\boldmath A Study of the Latest Updates of the Readout System for the Hybird-Pixel Detector at HEPS}

\author[*,a,b,c]{Hangxu. Li,\note{Corresponding author.}}
\author[b,c]{Jie. Zhang,}
\author[b,c]{Wei. Wei,}
\author[a,b,c]{Zhenjie. Li,}
\author[b,c]{Xiaolu. Ji,}
\author[a,b]{Yan. Zhang,}
\author[b,c]{Xuanzheng. Yang,}
\author[b,c]{Shuihan. Zhang,}
\author[d]{Xueke. Ma,}
\author[a,b,c]{Peng. Liu,}
\author[b,c]{Zheng. Wang,}
\author[b,c]{Yuanbai. Chen}

\affiliation[a]{Beijing Synchrotron Radiation Facility, High Energy Photon Source, Institute of High Energy Physics, Chinese Academy of Sciences,Beijing,China}
\affiliation[b]{University of Chinese Academy of Sciences,Beijing,China}
\affiliation[c]{State Key Laboratory of Particle Detection and Electronics, Institute of High Energy Physics, Chinese Academy of Sciences,Beijing,China}
\affiliation[d]{College of Electrical and Information Engineering, Shaanxi University of Science and Technology,Xi'an,China}
\emailAdd{lihx@ihep.ac.cn}

\abstract{The High Energy Photon Source (HEPS) represents a fourth-generation light source. This facility has made unprecedented advancements in accelerator technology, necessitating the development of new detectors to satisfy physical requirements such as single-photon resolution, large dynamic range, and high frame rates. Since 2016, the Institute of High Energy Physics has introduced the first user-experimental hybrid pixel detector, progressing to the fourth-generation million-pixel detector designed for challenging conditions, with the dual-threshold single-photon detector HEPS-Beijing PIXel (HEPS-BPIX) set as the next-generation target. HEPS-BPIX will employ the entirely new Application-Specific Integrated Circuit (ASIC) BP40 for pixel information readout. Data flow will be managed and controlled through readout electronics based on a two-tier Field-Programmable Gate Array (FPGA) system: the Front-End Electronics (FEE) and the Input-Output Board (IOB) handle the fan-out for 12 ASICs, and the $\mu$4FCP is tasked with processing serial data on high-speed links, transferring pixel-level data to the back-end RTM and $\mu$TCA chassis, or independently outputting through a network port, enabling remote control of the entire detector. The new HEPS-BPIX firmware has undergone a comprehensive redesign and update to meet the electronic characteristics of the new chip and to improve the overall performance of the detector. We provide an overview of the core subunits of HEPS-BPIX, emphasizing the readout system, evaluating the new hardware and firmware, and highlighting some of its innovative features and characteristics.}


\arxivnumber{1234.56789} 

\proceeding{N$^{\text{th}}$ Workshop on X\\
  when\\
  where}

\begin{document}
\maketitle
\flushbottom

\section{Introduction}
\label{sec:intro}

HEPS is one of the most advanced facilities located in Beijing, China, aimed at producing X-ray synchrotron radiation with unprecedented brightness. The high-brightness radiation of  $5\times 10^{22}$ photons $s^{-1}$ $mm^{-2}$ $mrad^{-2}$ $\left( 0.1\%bandwidth \right) ^{-1}$ allows scientists to analyze the internal structures of materials and molecules. Owing to the vast amounts of data generated by HEPS experiments, with data rates reaching several hundred Gb/s, the development of readout electronics must accommodate a variety of requirements, such as:
\begin{enumerate}
  \item The impact of the front-end module's dimensions and packaging on the yield rate.
  \item Data transmission and bandwidth of the front-end ASIC.
  \item The capability for online data processing to diminish the requirements for backend data transmission and storage.
  \item The scalability of integrating multiple modules to accommodate the larger detection areas of future detectors.
  \item Flexibility to facilitate upgrades with new algorithms.
\end{enumerate}

\par PC-based technology suffers from high latency and lower reliability in the first-level data acquisition for detectors; in contrast, FPGAs have significant advantages in terms of flexibility and development time. For these reasons, along with the specialized customization requirements of HEPS-BPIX, FPGAs are the best choice for detector data processing. Ultimately, the HEPS-BPIX is completed with a multi-level PCB developed around Field-Programmable Gate Arrays (FPGAs).
\par In this article, we introduce the improvements in the readout system over the previous generation's architecture, along with the development of firmware and several new features. The structure is as follows: Section II outlines the HEPS environment and the data processing architecture currently employed in HEPS; Sections III and IV cover the data readout at the front-end and back-end, respectively, focusing on the hardware and firmware structures as well as the improvements in the new release. Lastly, the preliminary test performance of the overall system is provided.

\section{HEPS-BPIX system}
\label{sec:intro}

The HEPS-BPIX is expected to be utilized at most of the beamlines in HEPS, particularly for large molecule crystallography, Laue diffraction, and other synchrotron radiation applications that demand noise-free pixels and high pixel resolution. Up to now, the HEPS-BPIX detector has been undergoing iterative upgrades and transitions from single-threshold to dual-threshold functionality\cite{a}\cite{b}\cite{c}\cite{d}\cite{e}. The system is required to handle a multitude of diverse types of information, and users have the capability to remotely connect to the detectors via independent software or a distributed control system. Consequently, we have developed all interfaces associated with the HEPS-BPIX detector and a standalone DAQ software based on C++, encompassing functionalities such as detector configuration, readout, and trigger acquisition.

\par As shown in figure 1, the 6-million-pixel detector of HEPS-BPIX is subdivided into 40 subunits. Subunit is an independently readable unit equipped with a complete front-end data acquisition system. The system consists of two FEEs (Front-End Electronics), one IOB (Input/Output Board), one $\mu$4FCP, and RTM (Rear Transition Module) electronics. Each subunit carries 12 BP40 chips, which amounts to 147456 dual-threshold pixels, with each threshold providing 14-bit information.

\par On the BP40, the signal generated by each pixel is first shaped by filters, then undergoes domain discrimination, and is output as a dual-threshold data stream through an LVDS serializer. In this scenario, considering a frame rate of 1 kHz, each module will generate a peak rate of 4.7 Gb/s ($256\times 576\times 32\times 1000\approx 4.71859\times 10^9\,\,bit/s$); Consequently, the final 6M prototype will produce a peak rate of 192 Gb/s ($6000000\times 32\times 1000\approx 1.92\times 10^{11}\,\,bit/s$).
\begin{figure}[htbp]
\centering 
\includegraphics[width=15cm]{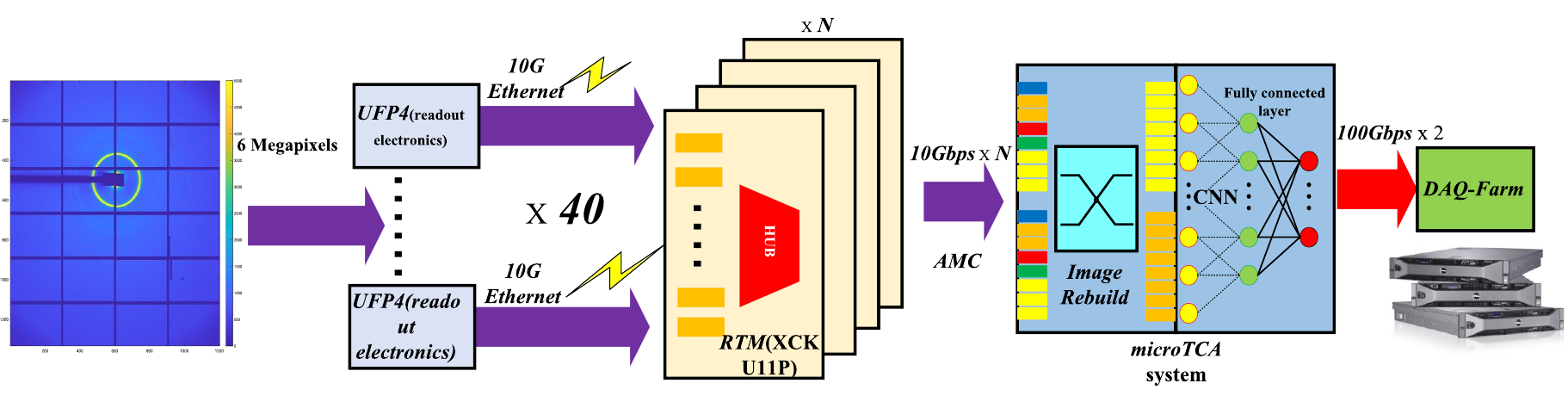}
\caption{\label{fig:i} The overall structure of the HEPS-BPIX with 6 million pixels}
\end{figure}

\par In the backend, each $\mu$4FCP is equipped with a powerful advanced mezzanine card (AMC) board, which provides a data link via Advanced Rear Transition Module Connectors (Zone 3) for the final aggregation and output of data at a bandwidth of 200 Gbps. The AMC board is loaded with a high-performance $Ultrascale^+$ FPGA, featuring up to 2700 DSP48 units for carrying out convolution and image processing\cite{f}\cite{g}. Considering that the 40 subunits have the same readout architecture, this paper will primarily focus on the architecture and performance of the subunit.

\section{Hardware}
\label{sec:intro}

The subunits are designed with a modular structure, where the Front-End Electronics (FEE) carrying ASICs and sensors are placed in the front-end cooling area for optimal noise performance; Meanwhile, the Input/Output Board (IOB) and $\mu$4FCP are housed in an $\mu$TCA chassis that is separate from the front-end, with flexible cables connecting the FEE and IOB. The following paragraphs will briefly outline the hardware associated with the readout system, which is categorized into several sections.

\subsection{FEE}

The FEE is the first-level readout of the subunit, as illustrated in figure 2; Its top layer is bump-bonded with twelve BP40s. The bottom layer abandons the backing board and extension board design of the previous generation, instead integrating direct-insert sockets and separate fan-out cables for all I/O (data, triggers, and control)\cite{e}.

\begin{figure}[htbp]
    \centering
    \subfloat[The encapsulated subunit and the FEE's PCB]{\label{fig:a}\includegraphics[width=4.5cm]{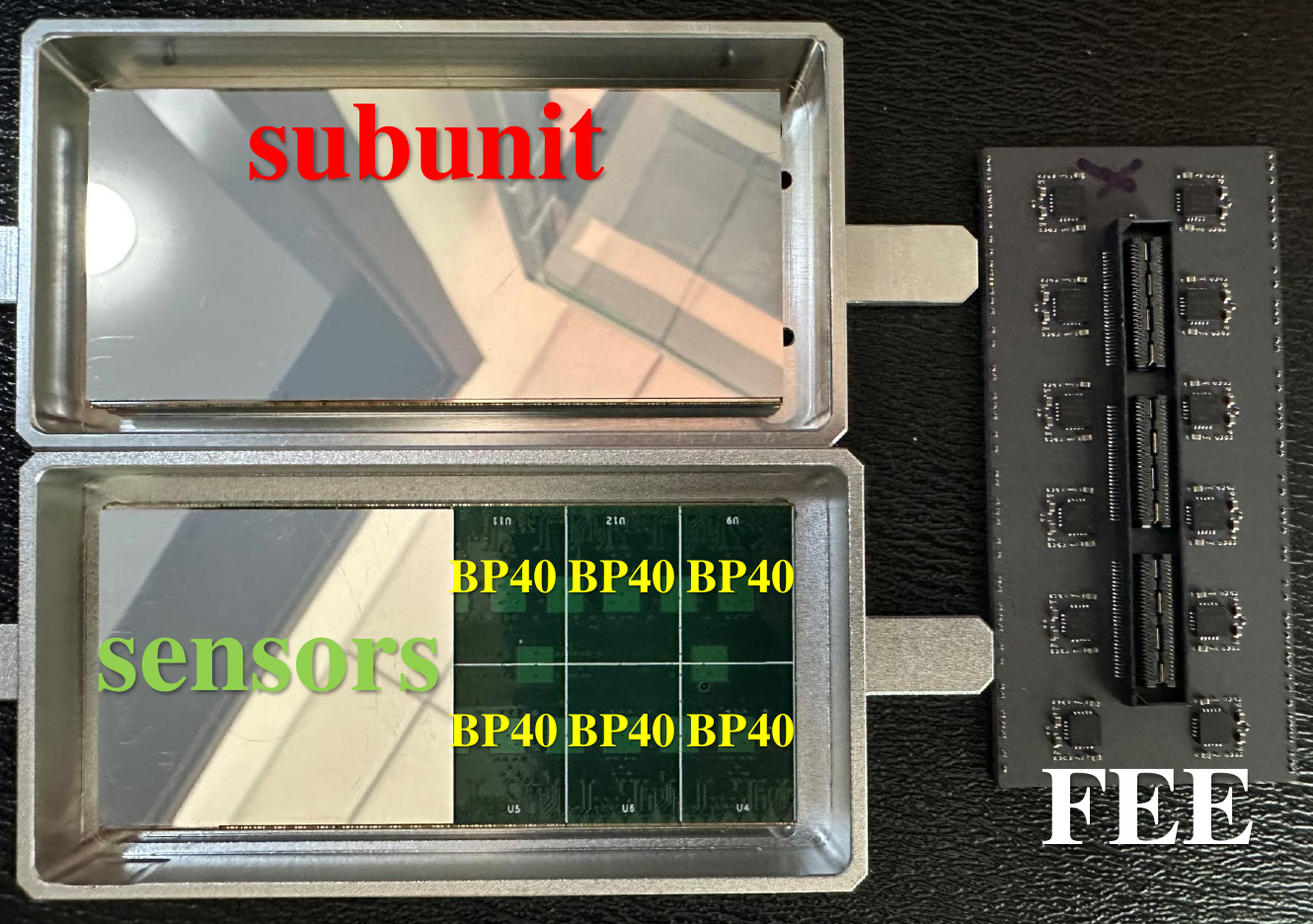}}
    \subfloat[The connection status among the 12 BP40s in the FEE]{\label{fig:b}\includegraphics[width=8cm]{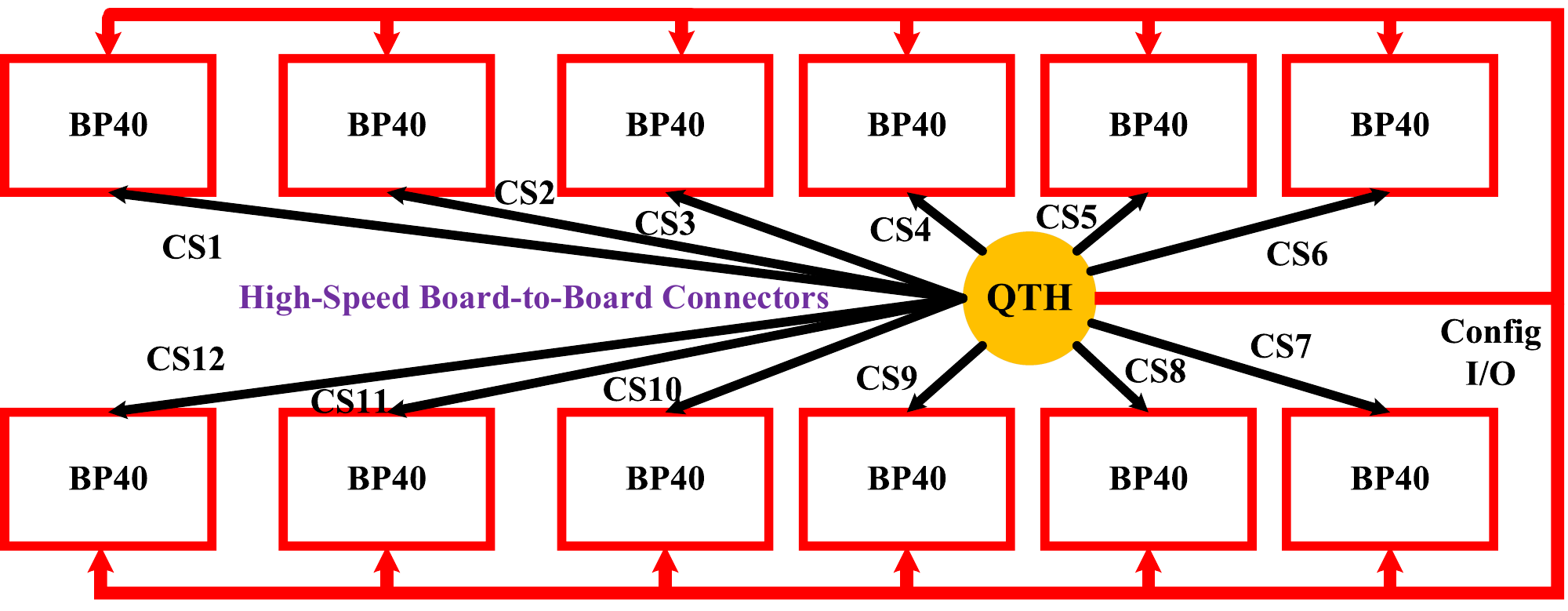}}
    \caption{Front-End Electronics}
    \label{fig:XXX.}
\end{figure}

\par Given the FEE's position at the front end, space constraints are extremely stringent. Compared to BPIX4, replacing multiple low-speed IOs on the readout bus with a single LVDS has effectively reduced the wire-bonding area between the ASIC and the FEE. This design significantly enhances the yield rate, thus substantially reducing costs while decreasing the stitching area between subunits. Additionally, considering the past issues of excessive thermal conductivity differences between FR4-PCBs and ASICs leading to heat accumulation and bump cracking, the FEE has been designed using ceramic materials.

\subsection{IOB}

The IOB receives IO signals fanned out from the FEE and redistributes them to the $\mu$4FCP for processing via the FMC. The figure below shows photographs of the IOB.

\begin{figure}[htbp]
    \centering
    \subfloat[TOP view]{\label{fig:a}\includegraphics[width=6.5cm]{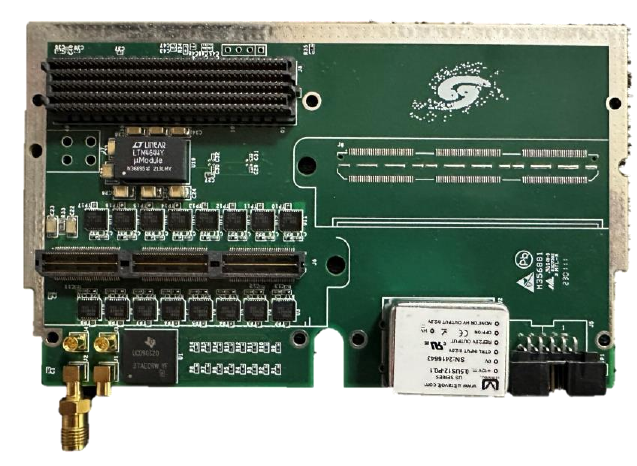}}
    \subfloat[BOTTOM view]{\label{fig:b}\includegraphics[width=6.8cm]{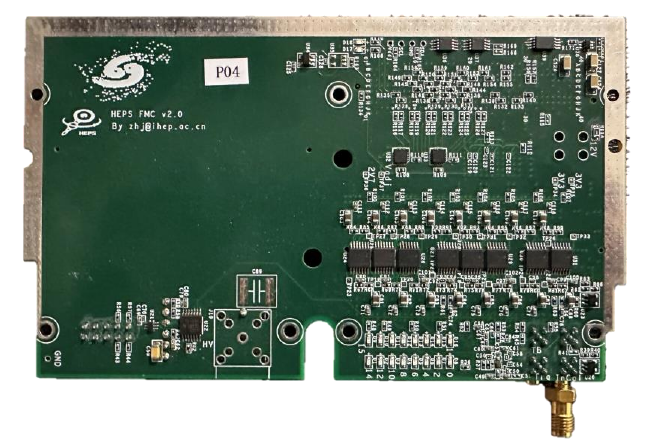}}
    \caption{IOB}
    \label{fig:XXX.}
\end{figure}

\par Based on previous experience, the temperature, voltage, and current of the detector determine whether the overall system is in a suitable operating state. The IOB's Low-Dropout Linear Regulator (LDO) output must travel through long cables to reach the load, and the impedance of the cables causes a voltage drop, resulting in a load voltage that is less than the ideal value. To ensure the voltage required by the BP40 is maintained under different load currents, the IOB can collect load voltage and current in real time through PMBUS and dynamically increase the LDO output to compensate for the voltage drop. Therefore, the IOB is designed to interconnect via the IIC bus on the hardware level. As shown in the figure 4, digital potentiometers controlled by IIC and UCDs controlled by PMBUS are all mounted on the bus as slave devices. For this purpose, an IO expander is integrated into the IOB to facilitate the control of 32 slave devices, allowing the FMC to monitor and control all environmental states with just two I/O  (SDA and SCL). Ethernet to Wishbone and Wishbone to I2C bus logic are integrated into the FPGA firmware. The LDO voltage and sensor high-voltage input-output curves demonstrate that an Ethernet-controlled digital potentiometer with 256-tap precision successfully functions like an 8-bit DAC with output currents at the amperes. Testing has revealed that integral nonlinearity can be achieved to within 1$\%$. It is possible to meet all the BP40's requirements for low voltage and high voltage (sensor high voltage 80 to 200 V), as well as the current specifications.
\begin{figure}[htbp]
    \centering
    \subfloat[IIC slave device connection diagram in IOB]{\label{fig:a}\includegraphics[width=7cm]{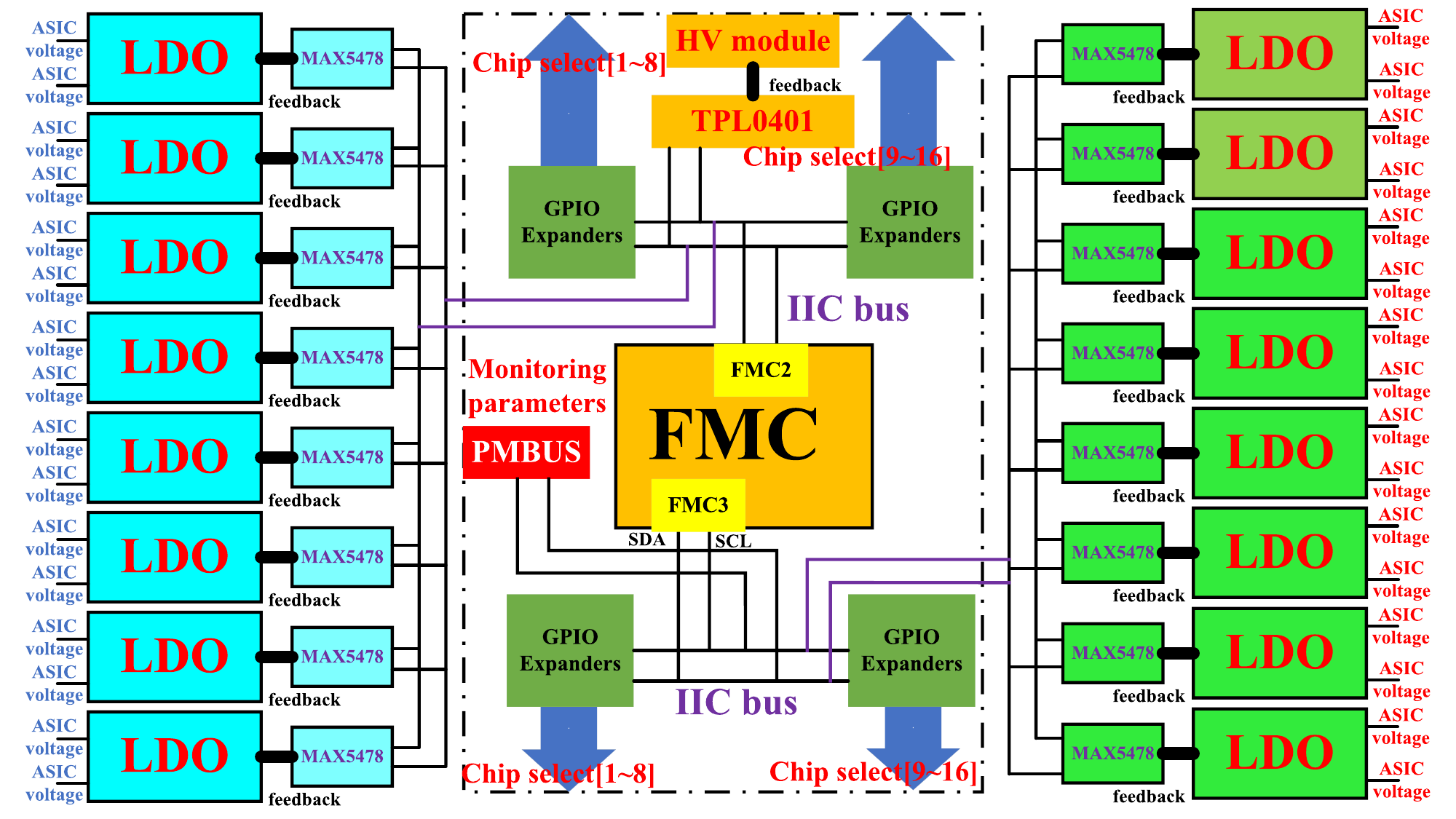}}
    \subfloat[The input-output relationship between the digital potentiometer, LDO voltage, and sensor high voltage]{\label{fig:b}\includegraphics[width=7.5cm]{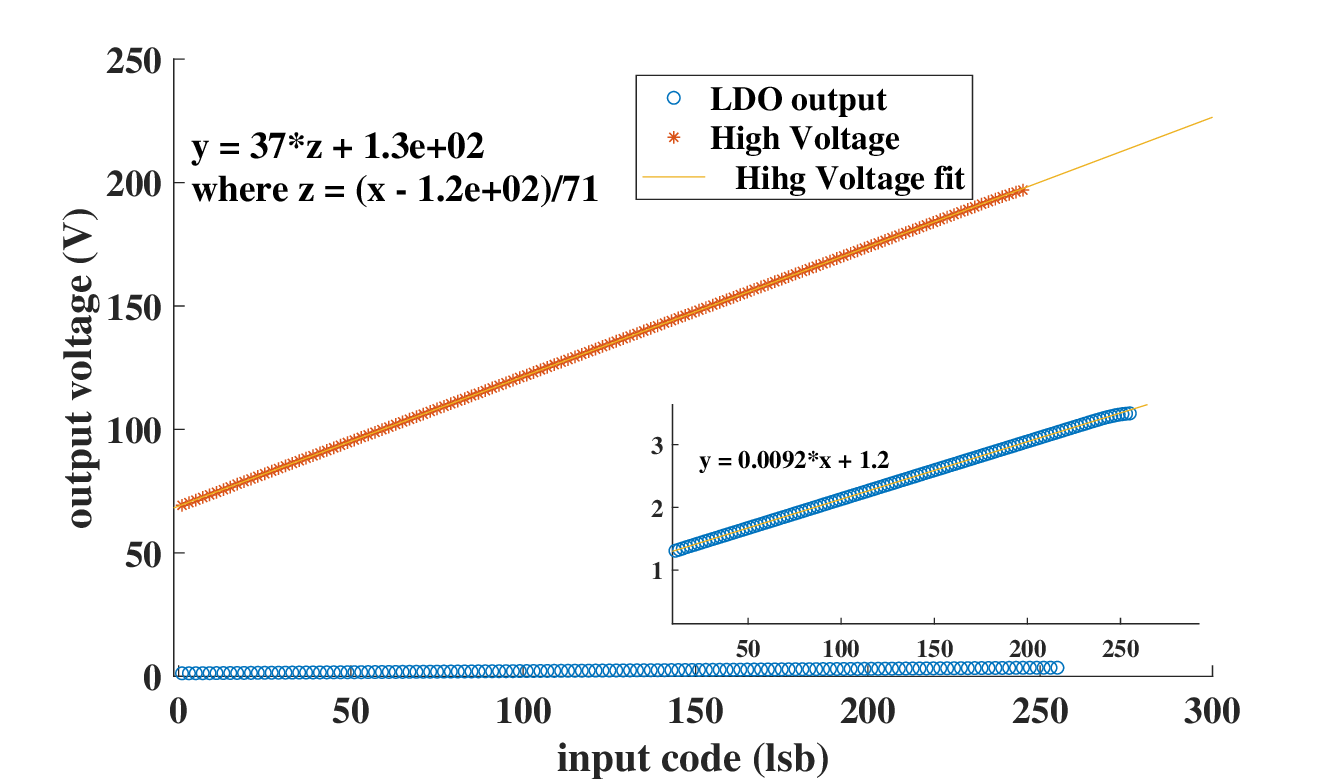}}
    \caption{The IIC structure of the IOB and related test results}
    \label{fig:XXX.}
\end{figure}

\subsection{$\mu$4FCP}

The $\mu$4FCP, equipped with the primary FPGA, reads out data from two FEE outputs, and after processing, the data stream is transmitted to the AMC board via 10G TCP/IP or directly to the DAQ through a gigabit RJ45 Ethernet port. Notably, the $\mu$4FCP can directly receive clock and control commands provided by the $\mu$TCA chassis, or from the onboard crystal oscillator and DAQ instructions. The figure 5 shows the subunit equipped with FEE, IOB, and $\mu$4FCP, where all logic and registers are provided by the FPGA. This part will be elaborated upon in the subsequent firmware section.

\begin{figure}[htbp]
\centering 
\includegraphics[width=13cm]{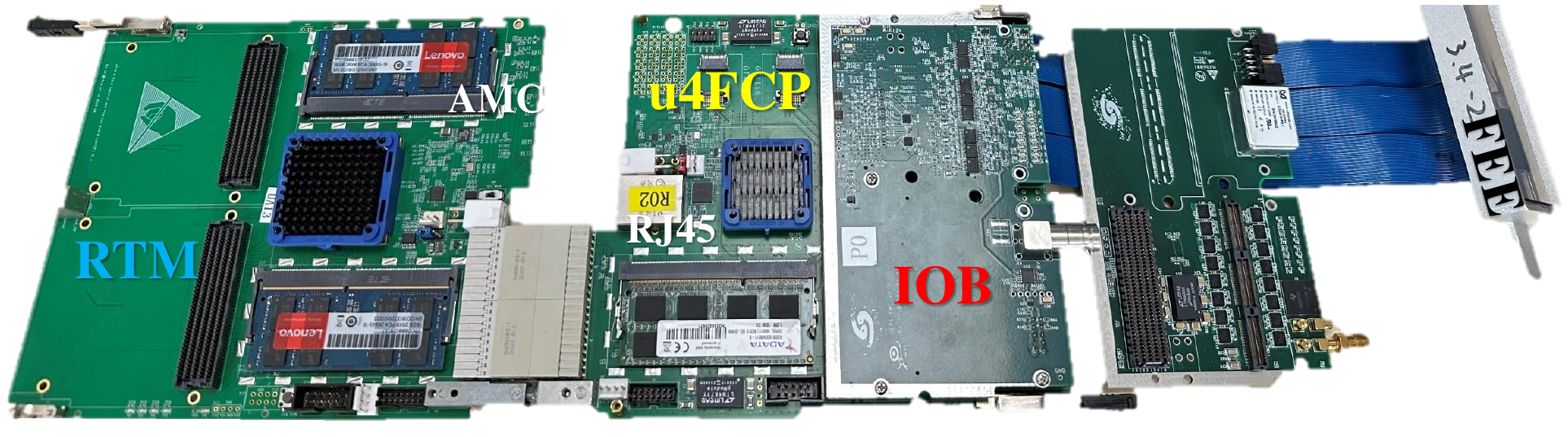}
\caption{\label{fig:i} The complete subunit after connecting FEE, IOB, and $\mu$4FCP}
\end{figure}

\section{Firmware}
\label{sec:intro}
The BP40 chip, featuring a novel architecture, requires firmware for configuration and readout, and due to chip and sensor PVT (Process, Voltage, and Temperature) variations, chips packaged within the same module require different parameters for calibration, to ensure a consistent response to X-rays. This section will provide a detailed description of the firmware designed for the $\mu$4FCP, with the block diagram of the firmware illustrated in the figure 6, and the firmware can be divided into the following parts for detailed discussion.

\begin{figure}[htbp]
\centering 
\includegraphics[width=10cm]{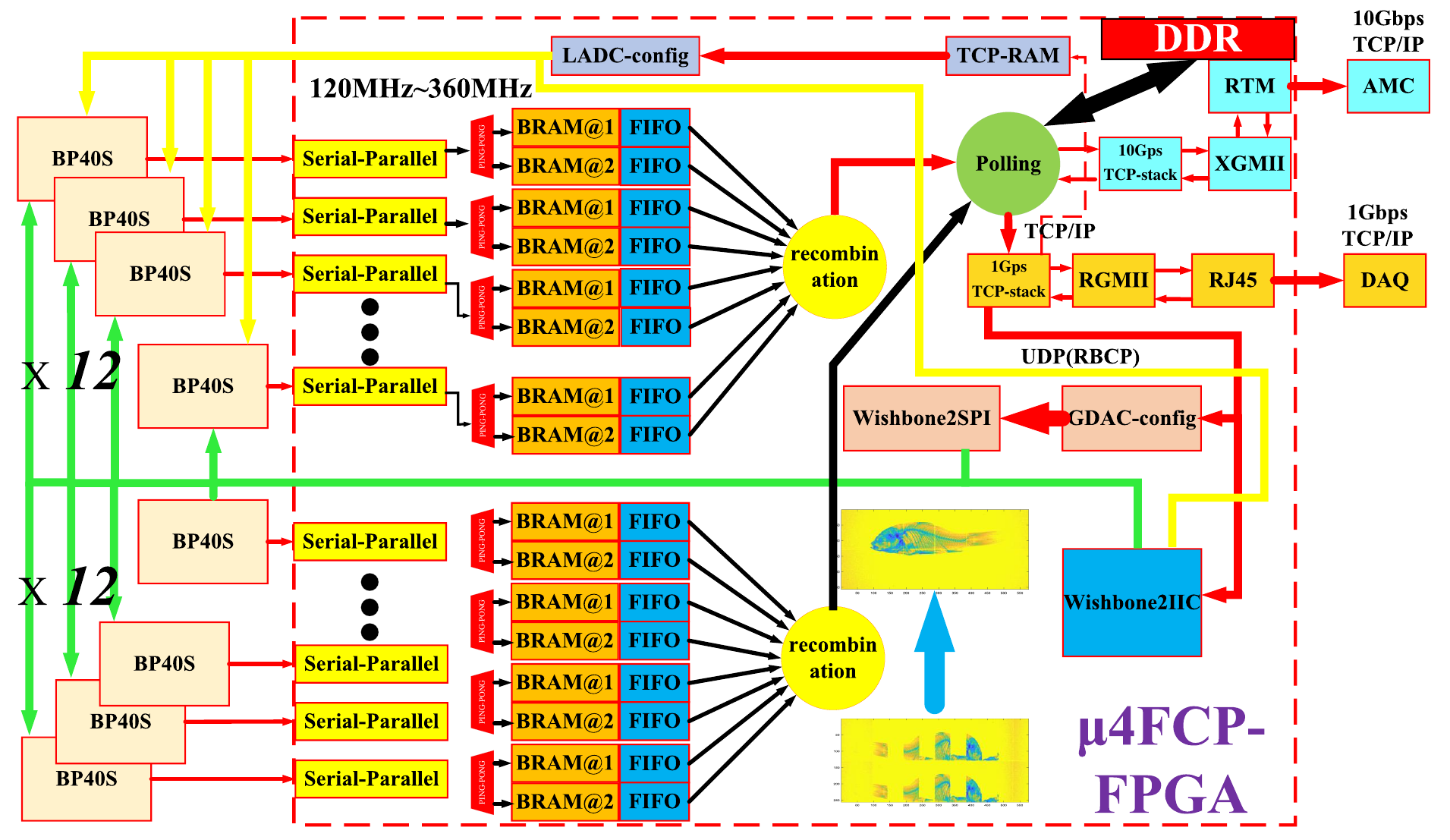}
\caption{\label{fig:i} Firmware structure of HEPS-BPIX40 subunit}
\end{figure}

\subsection{Global Parameter Configuration}

The BP40 has configurable global parameters, with up to 24 types of gain and offset parameters designed for different X-ray energies. To achieve flexible control over each chip, all parameters must be accessible for software control. Given the complex instruction configuration of previous DAQ versions, not all register addresses on the FPGA are allocated to the DAQ; instead, only a few essential register interfaces are made available, such as for frame rate settings, trigger settings, and start and stop commands, while other parameters are provided to the user through a streaming interface. For this streaming interface, the Wishbone bus is employed, which is an on-chip bus standard introduced by Silicore Corporation. This bus is characterized by its simplicity, flexibility, and openness, and is currently maintained by OpenCores\cite{h}. 

\begin{figure}[htbp]
\centering 
\includegraphics[width=10cm]{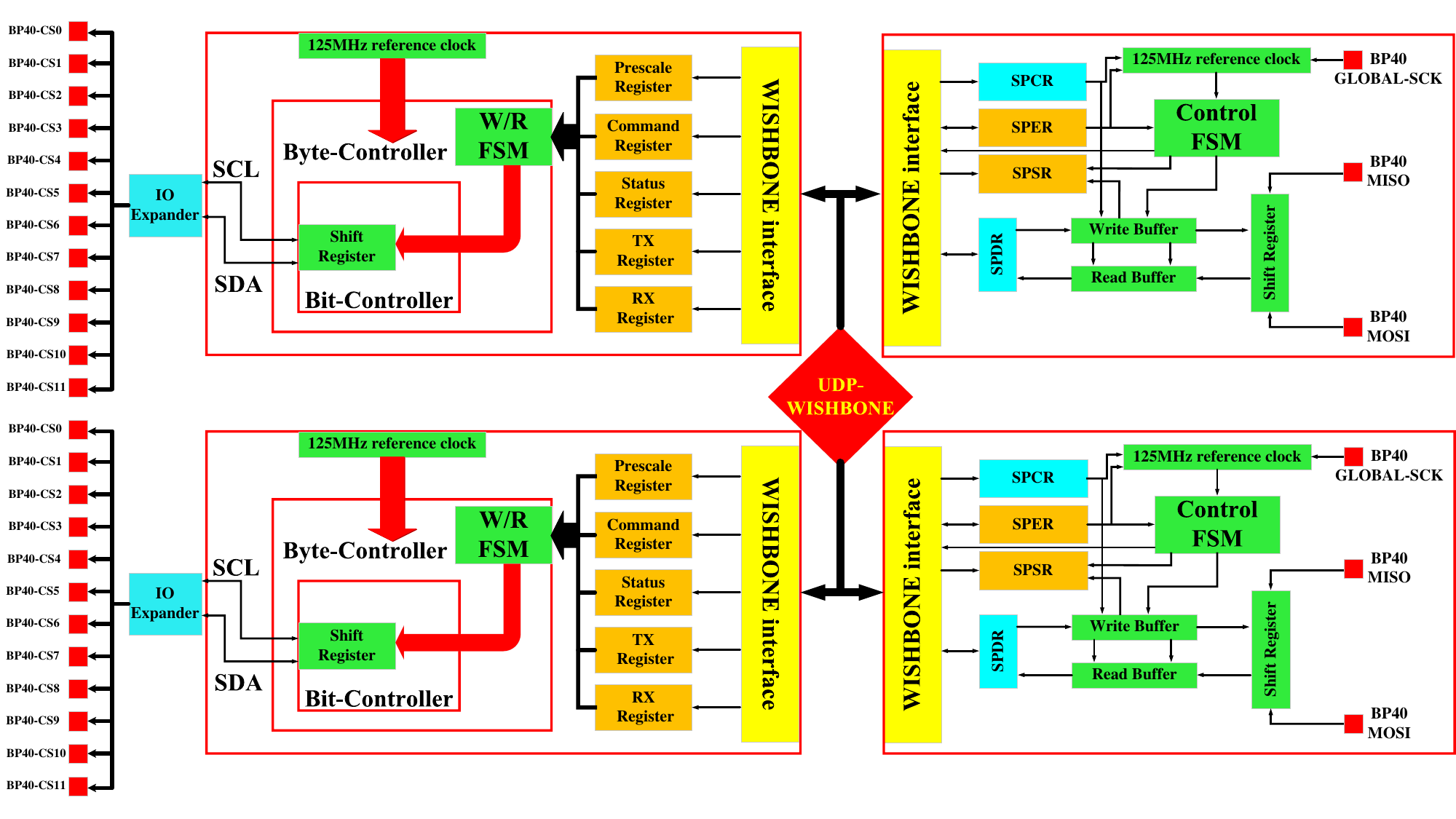}
\caption{\label{fig:i} The architecture of BP40's Global parameter configuration chain}
\end{figure}

\par As shown in figure 7, Within the firmware, multiple IIC cores and SPI cores are mounted on the standard Wishbone interface. Since the configuration of global parameters is similar to SPI, the data stream can be completed with just one register. The SPI clock is derived from a 125MHz reference clock through division, with parameters such as the division factor and phase determined by another register. Therefore, the DAQ can configure all global parameters using just two registers.

\subsection{Pixel-level Independent Parameter Configuration}
The pixels in the BP40 also have independent parameters, with each pixel comprising 28 bits (including two 5-bit local thresholds, a 2-bit preamplifier gain, a calibration input enable bit, and padding zero bits), totaling 147456*28 bits. To save on bond pad area, as shown in figure 8, the configuration data chains $ARRAYIN\left[ 3:0 \right] $ of the 12 BP40 chips are interconnected, with the chip select signal accomplished by CS (Chip Select). 

\begin{figure}[htbp]
\centering 
\includegraphics[width=12cm]{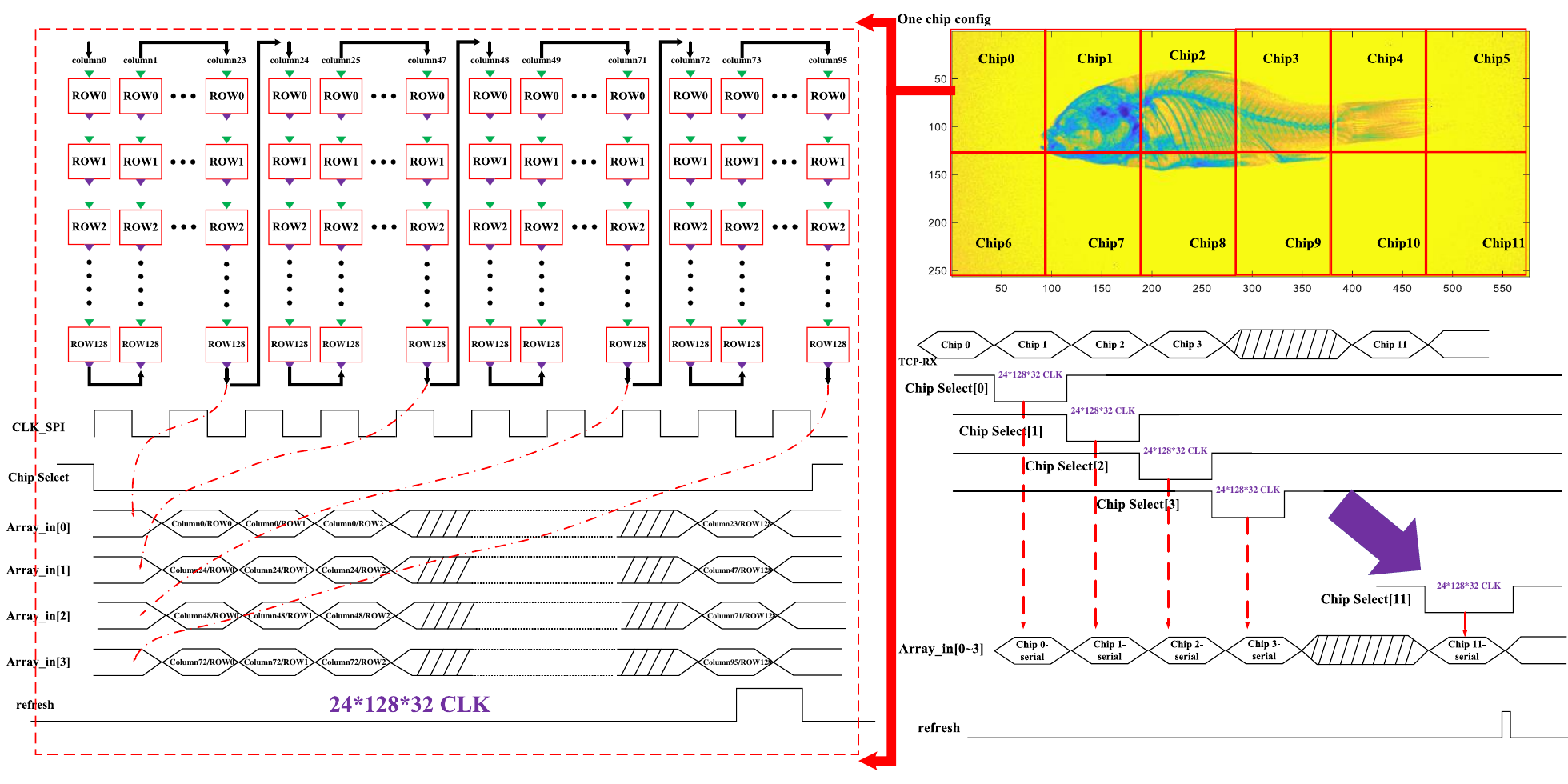}
\caption{\label{fig:i} The timing relationship of BP40's independent parameter configuration chain}
\end{figure}

Each chip is further divided into four regions, with pixels in adjacent regions connected end-to-end, and pixel information within a region is to be configured into the BP40 in a snake-like serial manner. Since the chip configuration depends on the CS signal, both standard configuration mode and calibration mode have been considered in the firmware design. In standard mode, each pixel requires an individually trimmed LDAC to standardize the response across all pixels, and the CS will time-share to complete the configuration of $ARRAYIN\left[ 3:0 \right] $. In calibration mode, the 12 chips pull down the CS simultaneously, standardizing the configuration parameters and reducing the configuration time by a factor of 12. From the perspective of FPGA buffer, the dual-threshold calibration tables for the two subunits require 9.4 Mb. The serial data chain also needs data reorganization and ping-pong buffering, and the 16 Mb (445*36 kb) of Kintex-7 inevitably lacks sufficient capacity. Therefore, to prevent this, we developed a streaming data interface from TCP to LDAC, synchronizing the TCP transmission to RAM with the configuration time of a single chip, pipelining the input to one-twelfth of the RAM. This approach allows for rapid refreshing of chip parameters while conserving the on-chip RAM of the FPGA.

\subsection{Pixel-level Readout}
After configuration, the frame interval of the BP40 can be determined either by registers in the software and internal FPGA counters or by an external trigger. When driven by the frame signal, the BP40 outputs the photon counts that have accumulated during the frame interval. A significant upgrade from the previous generation of chips is the shift from 72 low-speed buses to a single high-speed serial LVDS interface. As illustrated in figure 9, each BP40 chip is divided into 12 regions, and the pixels in each region serialize their output in a snake-like pattern through LVDS. The serialized data is not output in big-endian order but follows the regional chain $ARRAYOUT\left[ 11:0 \right] $, as depicted in the figure.

\begin{figure}[htbp]
\centering 
\includegraphics[width=12cm]{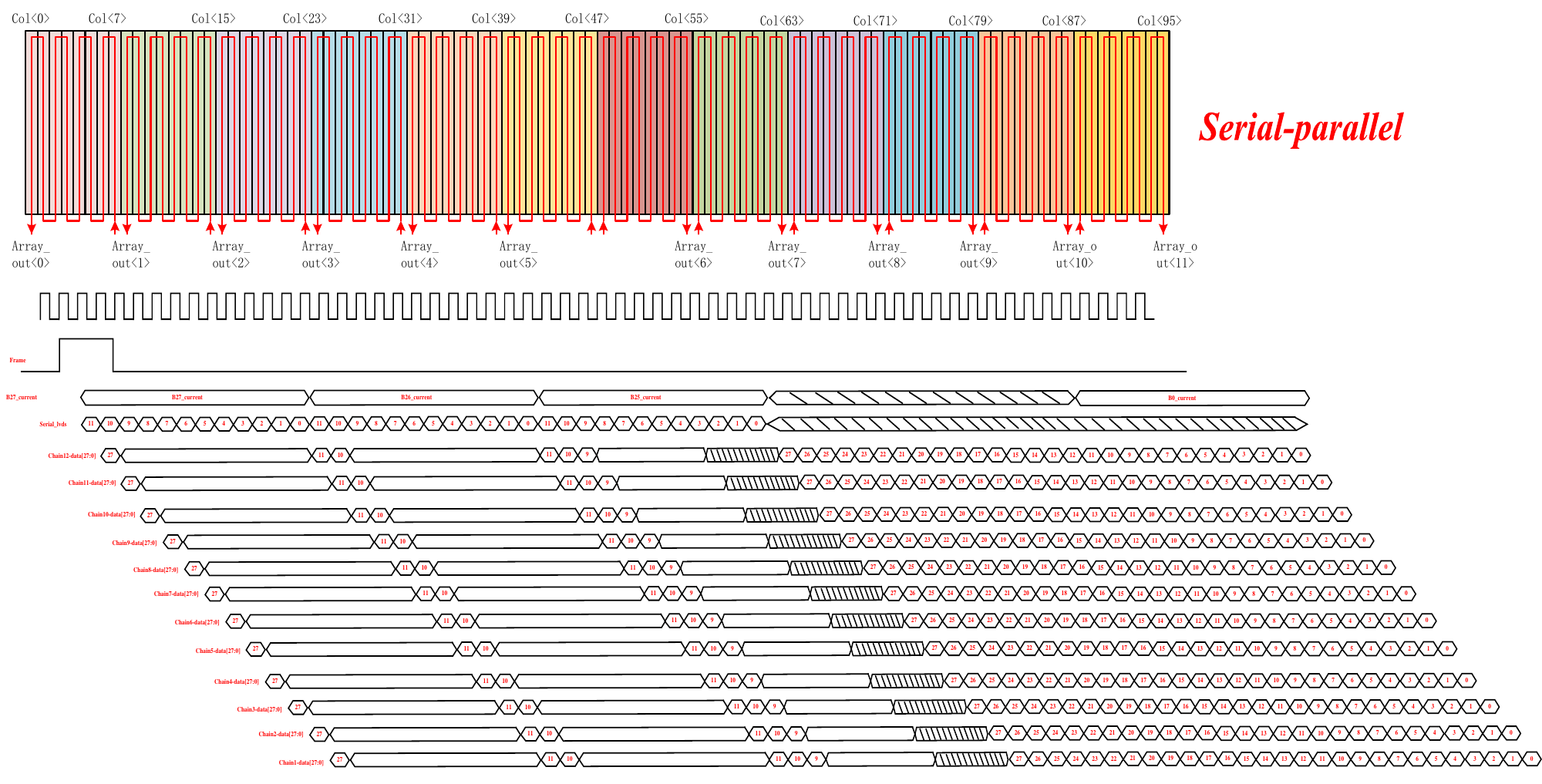}
\caption{\label{fig:i} Serial-to-parallel conversion in BP40's LVDS data link}
\end{figure}

\par Each pixel counts two thresholds, necessitating a 360 MHz serial LVDS frequency to support a frame rate of 1kHz. Notably, the data path goes through PCB routing and cables, but the sampling clock does not travel through the cables. Instead, it is generated directly on the FPGA, leading to potential data-clock desynchronization. To address this, input delay constraints are applied in VIVADO to compensate for the data path delay. Long-term testing has established that firmware with these constraints reads out data reliably.

\par After serial-to-parallel conversion, to prevent instability in TCP/IP transmission, an AXI4 to DDR3 MIG (Memory Interface Generator) data pathway is integrated within the FPGA, effectively turning the DDR3 into an 8GB FIFO (First-In, First-Out memory). The data retrieved from DDR3 is ultimately output to the DAQ (Data Acquisition) or AMC (Advanced Mezzanine Card) board via RJ45 or fiber optics. This integration ensures that even if there are fluctuations in TCP/IP transmission, the data can be buffered by the large capacity DDR3 memory before being sent out, thus providing a more stable data transfer process.

\section{Performance evaluation on prototype detector}
\label{sec:intro}
\subsection{Electronics performance}
Due to the detector's photon response being affected by noise, it is necessary to evaluate the overall system. The amplitude of the electrical pulses can be correlated with photon energy and is maintained at a fixed number of pulses to prevent photon pile-up effects, thereby enabling the stable retrieval of the charge response curve for each pixel. Given that the BP40 integrates a parasitic capacitance of 1.6 fF, a 250 mV pulse is equivalent to a photon of 8 keV. For this purpose, we developed a Matlab program for automated testing, which can control the AFG3252 via Ethernet to alter the amplitude of the step signal. Figure 10 displays the LDAC scan curves. The test reflects the pixel's response from 4.16 keV to 8 keV, and it can be demonstrated that the filter output of BP40 is linear.



\begin{figure}[htbp]
\centering 
\includegraphics[width=10cm]{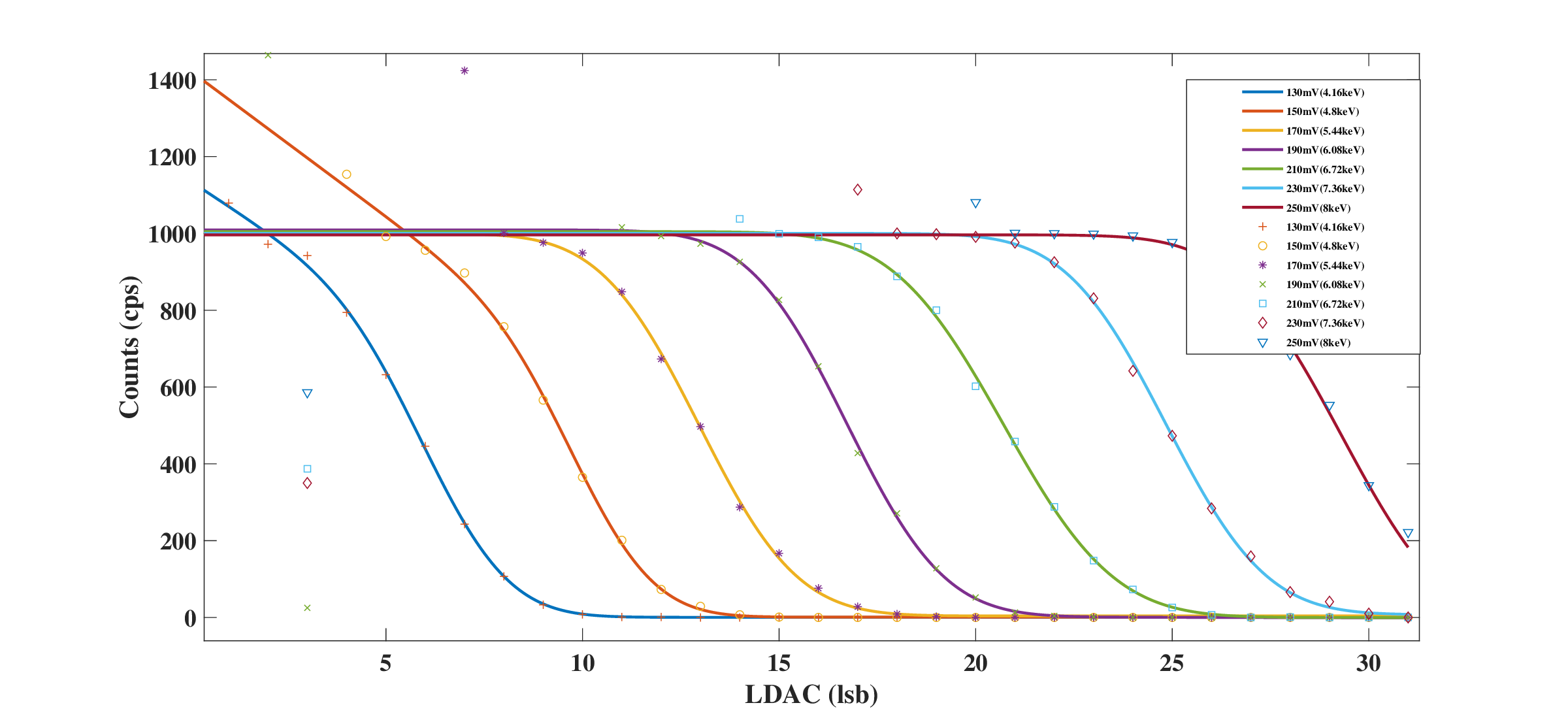}
\caption{\label{fig:i} LDAC scan curves}
\end{figure}


\par To accommodate photon conditions ranging from low to high energy, the BP40 offers four levels of gain. The GDAC is utilized to adjust the overall energy threshold of the detector system, thus it is necessary to define a relationship between GDAC, energy, and gain. Given that GDAC possesses a 10-bit range, it achieves greater precision compared to LDAC's Least Significant Bit (LSB). 
\par As shown in figure 11, with the aid of Arbexpress and MATLAB programs, an amplitude-modulated signal with a carrier frequency of 1 kHz was constructed as the input source, and the charge response curves of the GDAC at various gains were measured.  

\par The relationship between energy and counts in single-photon counting detectors conforms to a Gaussian error function and can be precisely fitted using the following equation\cite{i}\cite{j}:
\begin{equation}\label{eq1}
f\left( E_{TH} \right) =erf\left( \frac{E_0-E_{TH}}{\sigma} \right) \times \left( M-C_S\times \left( E_0-E_{TH} \right) \right) 
\end{equation}

\begin{equation}\label{eq1}
erf\left( x \right) =\frac{1}{\sigma \sqrt{2\pi}}\int_{-\infty}^X{e^{\frac{-\left( x-\mu \right) ^2}{2\sigma ^2}}dx}
\end{equation}

\par Herein, $M$ represents the mean flux, $\sigma$ denotes the gradient of the inflection point, $E_{TH}$ is the threshold corresponding to the inflection point of the curve, and $C_S$ is a slope related to charge sharing\cite{i}.

\par As shown in figure 12, after fitting and differentiating the average count values across all pixels, six energy points can be observed, and four gain equations have been derived. For the BP40, the GDAC test results at room temperature yield gains of 43.516 eV/LSB (Gain I), 34.06 eV/LSB (Gain II), 28.653 eV/LSB (Gain III),and 15.738 eV/LSB (Gain IV), respectively.


\begin{figure}[htbp]
	\centering
    {\includegraphics[width=10cm]{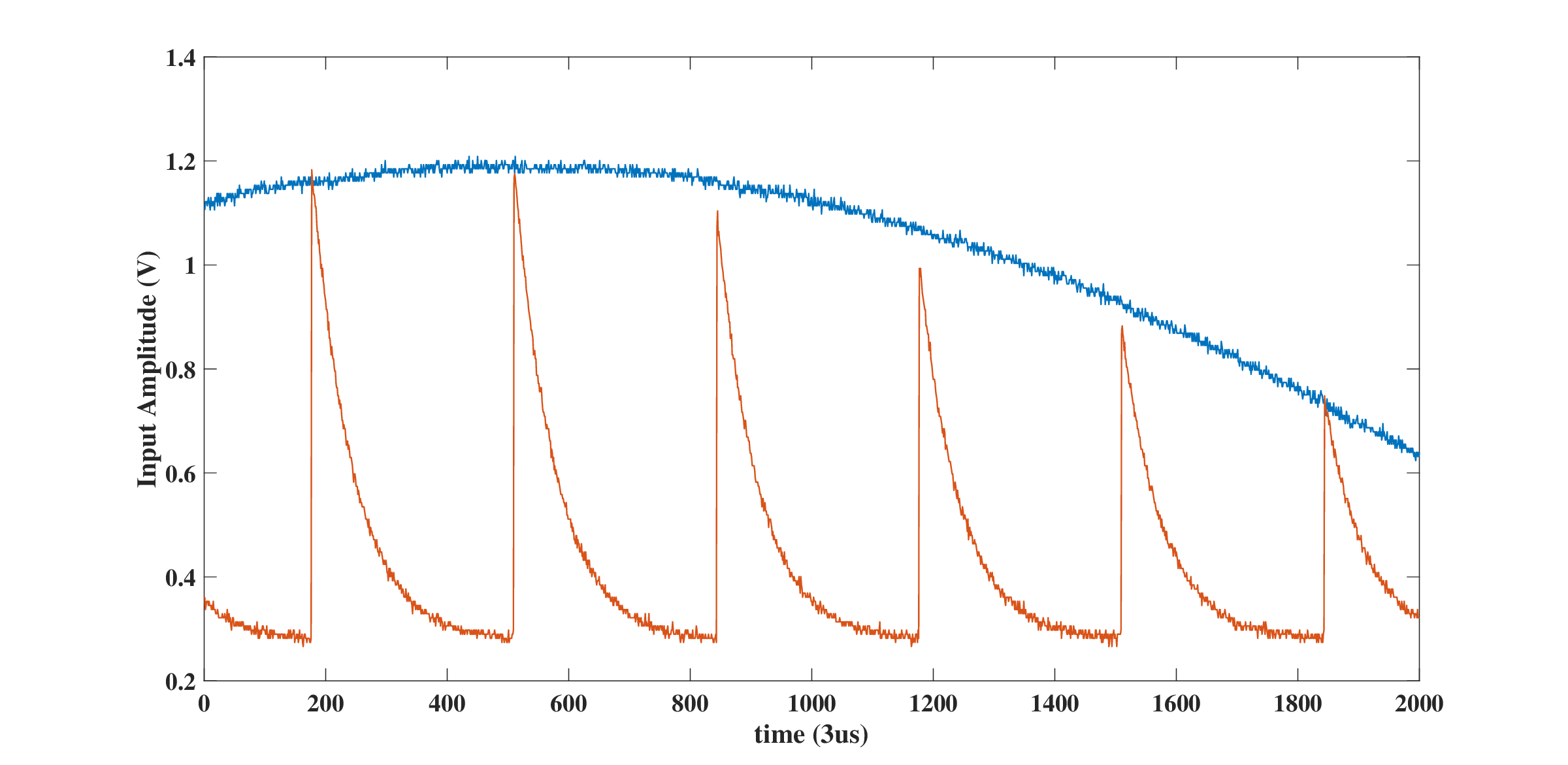}}
    \caption{An amplitude-modulated signal source with an input frequency of 1 kHz}
    \label{fig:XXX.}
\end{figure}

\begin{figure}[htbp]
	\centering
    \subfloat[Test curves of pixel count versus GDAC]{\label{fig:a}\includegraphics[width=7.5cm]{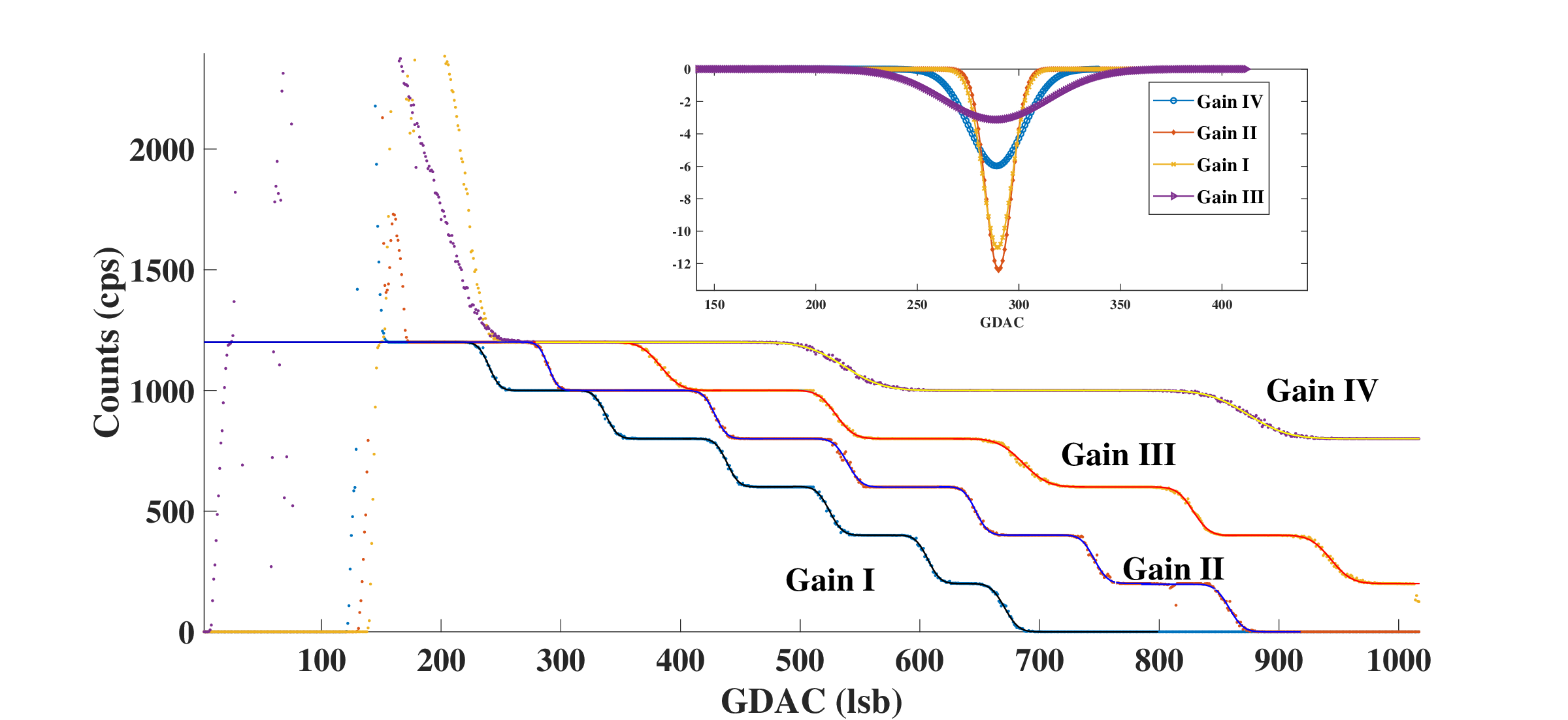}}
    \subfloat[The linear fitting curves between GDAC and various gains for the BP40]{\label{fig:b}\includegraphics[width=7.5cm]{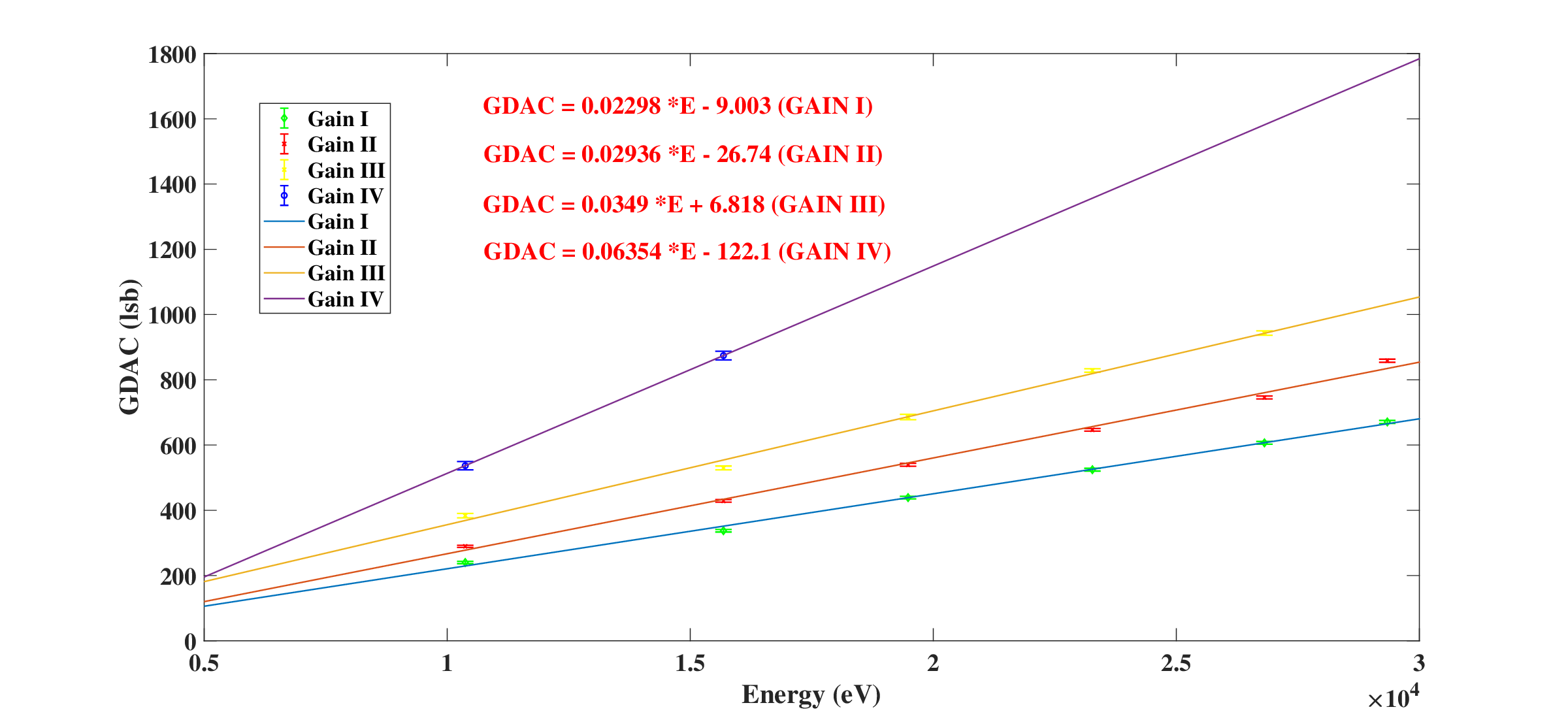}}
    \caption{The relationship between GDAC and pixel count under different gain conditions of BP40}
    \label{fig:XXX.}
\end{figure}


\subsection{Detector performance with X-ray}

	A detector readout prototype with a resolution of 140 micrometers per pixel and a 256*576 pixel array has been constructed. In the subunit, silicon-based sensors are bump-bonded to twelve BP40 chips, and the FEE (Front-End Electronics) and IOB (Input/Output Block) are connected via cables. figure 13 presents the imaging of peas in front of the subunit under the irradiation of a Au-anode X-ray tube, operating at 20 kV and 50 µA. To verify the overall performance of the dual-threshold system, the low threshold was set under 70 GDAC, and the high threshold at 300 GDAC. Concurrently, the BP40 operated in Gain II, with a trimmed LDAC configured to eliminate inconsistencies\cite{j}. Based on the GDAC-Gain curve, photons with $L_{\alpha}$, $L_{\beta}$, and $L_{\gamma}$ energies excited from the Au are shielded at 300 GDAC, whereas at 70 GDAC, most of the energy photons surpass the threshold.


\begin{figure}[htbp]
    \centering
    \subfloat[The low threshold is set to 70 GDAC]{\label{fig:a}\includegraphics[width=7.6cm]{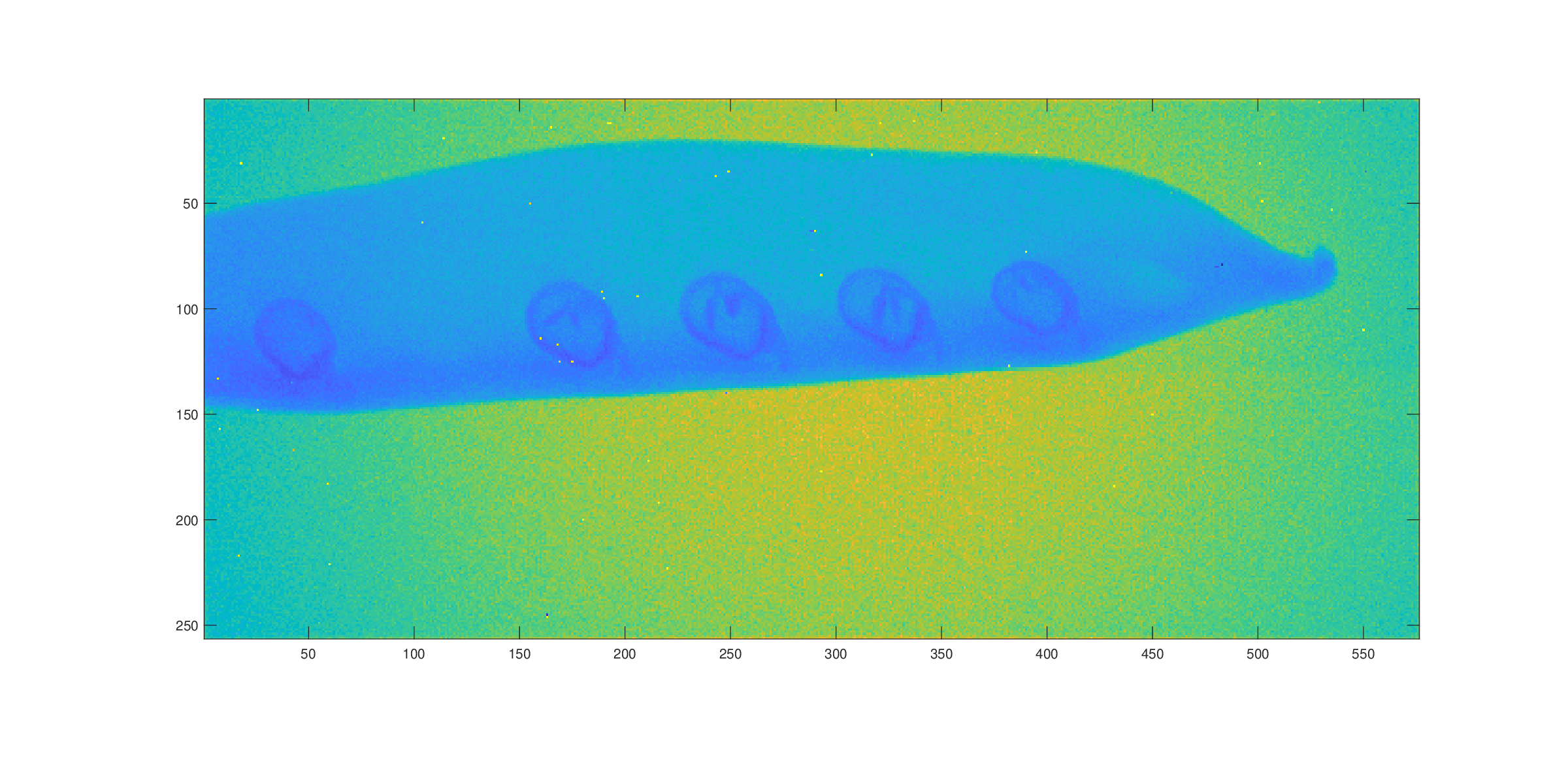}}
    \subfloat[The high threshold is set to 300 GDAC]
	{\label{fig:b}\includegraphics[width=7.6cm]{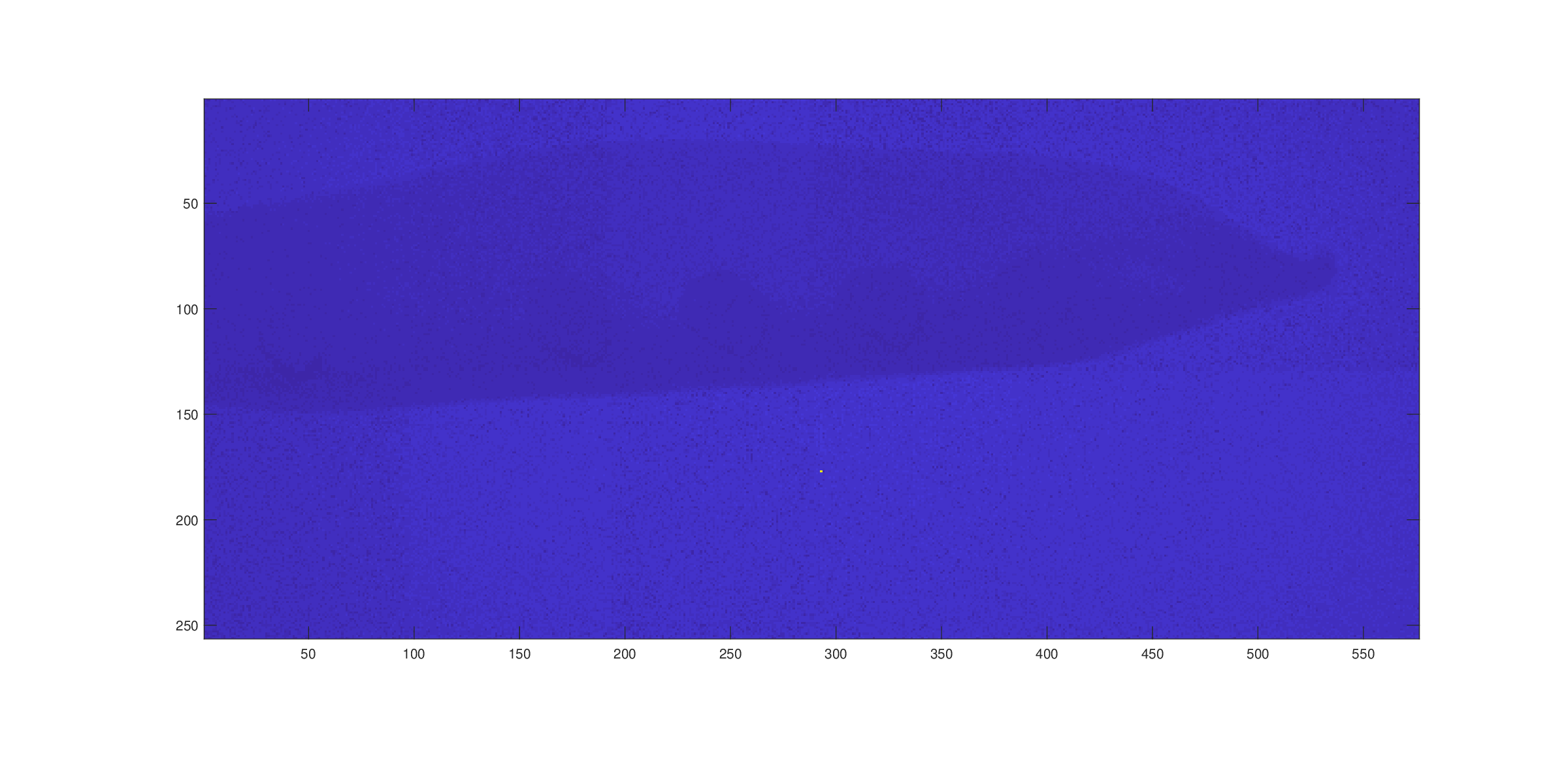}}
    \caption{Imaging of peas under X-ray exposure}
    \label{fig:XXX.}
\end{figure}

\par Finally, thanks to the monochromatic light provided by the BSRF light source at the 1W2B beamline, we have completed the performance testing of the detector under Monochromatic light. The detector response is uniform in the range of 6 keV to 20 keV, and the bad pixel rate is within 0.3$\%$. Preliminary tests indicate that the X-ray results are consistent with the electronics tests, proving that the hardware and firmware solutions have been successful. In our future work, we will employ $\mu$TCA chassis and several AMC boards to process the dual-threshold data for 6 million pixels, and we will complete the experiments under HEPS.

\section{Conclusion}
\label{sec:intro}

This article provides a concise introduction to the new HEPS-BPIX detector to be applied in the High Energy Photon Source (HEPS), offers a detailed description of its hardware and firmware functionalities, proposes diverse solutions, and ultimately confirms the functionalities. With this system, we report for the first time the performance of the BP40 and the sensor after packaging. The front-end is capable of continuous imaging at 1 kHz (under dual-threshold conditions) and can detect photon energies ranging from 6 to 20 keV at room temperature. Compared to the previous generation, the dead zone has been reduced and the yield rate has significantly improved. Furthermore, a fully adjustable firmware has been developed, which allows for the monitoring and regulation of all environmental parameters via the network and permits bias compensation for all chips.




\acknowledgments

This work was supported by the High Energy Photon Source (HEPS), a major national science and technology infrastructure in China, Platform of Advanced Photon Source Technology R\&D (PAPS) and the National Talent Program–Outstanding Young Talent (WRQB). We acknowledge the beam time granted at 1W2B as well as excellent technical support from the Beijing Synchrotron Radiation Facility and the BEPC accelerator.




\begin{thebibliography}{99}
\bibitem{a}
Wei Wei, Jie Zhang, Zhe Ning, Yunpeng Lu, \emph{HEPS-BPIX, a single photon counting pixel detector with a high frame rate for the HEPS project},Nuclear Instruments and Methods in Physics Research Section A: Accelerators, Spectrometers, Detectors and Associated Equipment (2016).
\bibitem{b}
Jie Zhang, Wei Wei, Jingzi Gu, Wei Shen, \emph{HEPS-BPIX, a hybrid pixel detector system for the High Energy Photon Source in China},Journal of Instrumentation (2017).
\bibitem{c}
Jie Zhang, Hangxu Li, Wei Wei, Jingzi Gu, Zhenjie Li, \emph{HEPS-BPIX2: The hybrid pixel detector upgrade for high energy photon source in China},Nuclear Instruments and Methods in Physics Research Section A: Accelerators, Spectrometers, Detectors and Associated Equipment (2020).
\bibitem{d}
Jie Zhang, Ye Ding, Wei Wei, Hangxu Li, Zhenjie Li, \emph{The TSV process in the hybrid pixel detector for the High Energy Photon Source},Nuclear Instruments and Methods in Physics Research Section A: Accelerators, Spectrometers, Detectors and Associated Equipment (2020).
\bibitem{e}
Hangxu Li, Jie Zhang, Ye Ding, Wei Wei, Zhenjie Li, \emph{A Dual Module Parallel Readout System Based on 10 Gb TCP/IP Transmission for HEPS-BPIX Detector},IEEE Transactions on Nuclear Science (2021).
\bibitem{f}
Jie Zhang, Hangxu L, Xiaoshan Jiang, \emph{The MicroTCA Fast Control Board for Generic Control and Data Acquisition Applications in HEP Experiments},IEEE Transactions on Nuclear Science (2018).
\bibitem{g}
Jie Zhang, Cong He, Aoqi Su, Manhao Qu, Yunhua Sun, Wei Wei, \emph{The MicroTCA.4 Fast Control and Processing Board for Generic Control and Data Acquisition Applications in HEP Experiments},IEEE Transactions on Nuclear Science (2023).
\bibitem{h}
Richard Herveille, \emph{WISHBONE System-on-Chip (SoC) Interconnection Architecture for Portable IP Cores.},Revision: B. (2002).
\bibitem{i}
S. Elbracht-Leong, A. Bergamaschi, D. Greiffenberg, D. Peake, R. Rassool, B. Schmitt, H. Toyokawa, B. Sobbott, \emph{Characterisation of an electron collecting CdTe strip sensor using the MYTHEN readout chip},Journal of Instrumentation (2015).
\bibitem{j}
Ye Ding, Zhenjie Li, Wei Wei, Jie Zhang, Hangxu Li, \emph{The Study of Calibration for the Hybrid Pixel Detector With Single Photon Counting in HEPS-BPIX},IEEE Transactions on Nuclear Science (2021).
\end{thebibliography}

\end{document}